\documentclass[11pt,a4paper]{article}
\usepackage{jheppub} 
\usepackage{graphicx}
\usepackage{bm}

\usepackage{caption}
\usepackage{framed}
\usepackage{xcolor}
\usepackage{rotating}
\usepackage{empheq}
\usepackage{afterpage}

\usepackage[utf8]{inputenc}
\title{\boldmath Recursive Generation of The Semi-Classical Expansion in Arbitrary Dimension}

\author{Cihan Pazarba{\c{s}}\i}
\affiliation{Physics Department, Bo\u{g}azi\c{c}i University\\
	34342 Bebek / Istanbul, Turkey}

\emailAdd{cihan.pazarbasi@boun.edu.tr}

\def\mrmd{{\mathrm{d}}}

\def\th{{\mathrm{th}}}

\def\exp{\mathrm{exp}}

\def\a{\alpha}

\def\G{\Gamma}
\def\d{\delta}

\def\ve{\varepsilon}
\def\D{\Delta}

\def\l{\lambda}
\def\L{\Lambda}

\def\z{\zeta}

\def\tU{\tilde{U}}

\def\mbfp{{\mathbf{p}}}
\def\mbfx{{\mathbf{x}}}

\def\mcalT{{\mathcal{T}}}

\def\msfT{\mathsf{T}}
\def\msfV{\mathsf{V}}
\def\msfH{\mathsf{H}}

\def\msfb{\mathsf{b}}
\def\msfc{\mathsf{c}}

\def\Tr{{\mathrm{Tr}}}

\def\<{\langle}
\def\>{\rangle}

\newcommand{\lbar}{\lower0.2ex\hbox{$\mathchar'26$}\mkern-10mu \lambda}

\usepackage{titling}

\begin{document}

	\begin{titlepage}

		\begin{center}{\Large \bf \boldmath Recursive Generation of The Semi-Classical Expansion in Arbitrary Dimension} 
		\end{center}
		\hrule
				\vskip 1cm

		\centerline{ {{\bf Cihan Pazarbaşı}}}

		\begin{center}
			\textit{Physics Department, Boğaziçi University\\
			34342 Bebek / Istanbul, TURKEY}
			
			\vskip 0.5cm 
			\texttt{cihan.pazarbasi@boun.edu.tr}\\

		\end{center}
		\vskip 1.0cm
		
		\centerline{\bf Abstract} \vskip 0.2cm \noindent 
		We present a recursive procedure, which is based on the small time expansion of the propagator, in order to generate a semi-classical expansion of the \textit{quantum action} for a quantum mechanical potential in arbitrary dimensions. In the method we use the spectral information emerges from the singularities of the propagator on the complex $t$ plane, which are handled by the $i\ve$ prescription and basic complex analysis. This feature allows for generalization to higher dimensions. We illustrate the procedure by providing simple examples in non-relativistic quantum mechanics.		
	
	\end{titlepage}
	\hrule
	\tableofcontents
	\vskip 0.5 cm
	\hrule
	\vskip 1.5 cm
\section{Introduction}
One of the main practical tasks of a theoretical physicist is constructing and solving differential or integral equations. However, it is a well-known fact that exact solutions to those equations are rare and perturbative methods are commonly used in many problems. Quantum theories are no different: Perturbation methods are at the center of practical calculations to obtain quantum observables.\\
\\
An important perturbative approach in quantum theories is the semi-classical approximation\footnote{Throughout this paper, the semi-classical expansion refers to the perturbative series in $\hbar$ while in the literature the semi-classical expansion sometimes refers to the terms of $e^{-1/\hbar}$. In fact, the exponentially suppressed terms, which generally have its own perturbative corrections, correspond to the non-perturbative part of the semi-classical expansion and together with the perturbative series, they form the trans-series structure of the semi-classical expansion. (See \cite{Aniceto:2018bis} for a comprehensive discussion on the subject.)}, which assumes $\hbar$ as its expansion parameter. It is also a tool to make a connection between the quantized theory and the underlying classical system. The relation between the quantized spectrum and the classical equations of motion, known as Bohr-Sommerfeld quantization, is older than the quantum theory we know today. Later, this relation was systematically used in non-relativistic quantum mechanics via the WKB method. However, soon the limitation of Bohr-Sommerfeld quantization in the presence of chaotic classical motion was realized. (See \cite{percival1977semiclassical} for a detailed review on the subject.) While this led to the development of EBK quantization and quantum chaos theory\footnote{For a nice book on the subject, see \cite{brack1997semiclassical}.}, it also implies a limitation of WKB theory itself. From the practical point of view, we can state that despite its many successful applications, WKB theory remains limited to effectively one dimensional systems.\\
\\
Another useful method to investigate the classical limit and its quantum corrections is the path integral. An advantage of the path integral method is its adaptability to both relativistic and non-relativistic theories in arbitrary dimensions, suggesting a unified picture for quantum theories. However, generating perturbative corrections turns into a computational burden very quickly and computing higher order corrections becomes practically impossible. This limits the applicability of this beautiful approach. \\
\\
At this point, it is reasonable to ask for the motivation behind computing higher order corrections. Why would only the first few corrections not be enough? An important reason is hidden in the resurgent structure of perturbative expansions \cite{Aniceto:2018bis}. Resurgence theory states that in most of the cases, naive perturbation theory leads to a divergent series and, on its own, only provides a partial answer. A more careful investigation leads to the emergence of non-perturbative effects hidden in the large order behaviour of the divergent expansion, and this is proposed as a possible way to construct complete solutions. \\
\\
Computation of quantum corrections based on the WKB approximation is a well studied subject. In fact, for one dimensional anharmonic oscillators, it has been automatized \cite{Sulejmanpasic:2016fwr}, based on the famous Bender-Wu approach \cite{Bender:1973rz}. Similar calculations have been done other on methods based on the geometric properties of the curve represented by the Hamiltonian of the system. One of them is based on the Picard-Fuchs differential equations, whose solutions correspond to the classical actions of the system \cite{Basar_2017}. Utilizing the recursion relation, the corresponding differential equation for the quantum corrections were also generated\footnote{Note that the generation of the quantum corrections via Picard-Fuchs equation are originally done for genus-1 potentials, while some extension to higher genus potentials are also recently discussed \cite{Raman:2020sgw}.} \cite{Mironov_2010,Kreshchuk:2018qpf,Fischbach:2018yiu,Raman:2020sgw}. Another method is based on the recursive nature of the holomorphic anomaly equations of topological string theory  \cite{Codesido_2017,Codesido:2017jwp}, which is related to the one dimensional quantum mechanics in the Nekrasov-Shatashvili limit \cite{NEKRASOV_2010}. All of these approaches utilize the recursive dependence of the quantum corrections to the classical term, i.e., $\hbar^0$ order. This is akin to topological recursion \cite{Eynard:2007kz,Norbury:2015lcn}, and it allows a systematic generation of quantum corrections with minimal input.\\
\\
As we mentioned above, all these analyses are done for one dimensional non-relativistic problems. The main motivation of this paper is tackling this problem by constructing a general procedure for higher dimensional quantum mechanical problems that can possibly be generalized to many body theories and quantum field theories. For these purposes, the investigation of the time translation operator will be at the center of this paper. The method is equivalent to a perturbative expansion via path integrals and this similarity supports our main objective. \\
\\
Before examining the propagator itself, in Section \ref{Section: SpectralProblem}, we start with its relation to spectral functions  and derive an integral representation of the so-called WKB action. Then, in Section \ref{Section: ExpansionArbitraryDimension}, which is the main section of this paper, we discuss the perturbative expansion of the propagator by utilizing a small time expansion and derive the recursion relation we were looking for. Note that at first, the time dependent formulation might appear to have a disadvantage for practical purposes despite its applicability to higher dimensional problems. However, by the construction we describe in Section \ref{Section: SpectralProblem}, in our approach the spectral information stems from the singularities of the time propagator on the complex time plane. This reduces the main computational task to the integration of ordinary integrals and basic complex analysis. In Section \ref{Section: AHO}, we will apply our method to anharmonic oscillators in arbitrary dimensions. The numerical results for this part are presented in Appendix \ref{Section: NumericalResults}. Finally, in Section \ref{Section: Discussion}, we finish the paper with a discussion of our analysis and an outlook to future work.

\section{Spectral Problem}\label{Section: SpectralProblem}
In this section, we will briefly review the spectral problem of a Hermitian  operator $\msfH$ acting on a Hilbert space. From elementary linear algebra, we know that the spectrum of $\msfH$ is given by the zeroes of the Fredholm determinant, i.e.,
\begin{equation}
D(u) = \det(u - \msfH) = 0 ,
\end{equation}
where $u$ represents the elements of the spectrum. Instead of dealing with $D(u)$ directly, we focus on another spectral function, which is the \textit{quantum action},\footnote{Our choice of labeling $\G$ as the \textit{quantum action} comes from the usage of the Fredholm determinant in QFT and many-body theories to calculate the effective actions (See, e.g., \cite{DUNNE_2005}). Its relation with the spectral $\z$ and $\Theta$ functions are also well-known \cite{voros1992}.} 
\begin{equation}\label{Action_definition}
\G(u) = \ln \det (u - \msfH) = \Tr \ln (u -\msfH).
\end{equation}
Now, the branch point of this new function carries the spectral information. One way to handle the singularity is introducing the resolvent $G(u) = (u - \msfH)^{-1}$ as
\begin{equation}\label{ActionFromResolvent}
\G(u)  = \int_{u_0}^u \mrmd z\,\Tr\, G(z),
\end{equation}
where $u_0$ is an arbitrary regular point\footnote{In another perspective, $u_0$ can be introduced to eliminate the infinities by a redefinition of the action as $\G_{\mathrm{fin}} =  \G(u) - \G(u_0)$. However, in our construction, we only encounter with the singularities that have a \textit{physical} meaning so there is no need for any regularization procedure.} of $G(z)$ on the complex $z$ plane. Note that the simple poles of $G(z)$, where the (discrete) spectrum  appears, correspond to the branch points of $\G(u)$ as demanded by construction \cite{Eden:1966dnq}. The information around the branch point can be obtained by employing the $i \ve$ prescription \cite[Appendix A]{Zinn_Justin_2004} and defining a gap for the action $\G(u)$ as 
\begin{align}\label{ActionGap_Defintion}
\D \G(u) = \G^+(u) - \G^-(u),
\end{align}
where we defined the actions in different branches as
\begin{equation} \label{Action_AnalyticalContinuation1}
\G^\pm(u) = \lim_{\ve\rightarrow 0 } \G(u\pm i\ve) = \lim_{\ve\rightarrow 0}\left\{\int_{0\pm i\ve}^{u\pm i\ve} \mrmd z\, \Tr \, G(z) \right\} .
\end{equation} 
It is well-known that the resolvent approach connects the classical dynamics and the quantum spectrum of $\msfH$ \cite{1970JMP....11.1791G,1971JMP....12..343G}. However, in perturbative calculations, it may become impractical. For this reason, it is more convenient to introduce its Fourier integral representation
\begin{equation}\label{ResolventToSchwinger}
G(u) = \pm i \int_{0}^\infty \mrmd t\, e^{\pm i (u - \msfH) t},
\end{equation}
where $t$ corresponds to a flow-time parameter conjugate to the eigenvalue $u$.\\
\\
Near the branch cut, $\G^\pm(u)$ becomes
\begin{equation}\label{Action_AnalyticalContinuation2}
\G^\pm(u) = - \lim_{\ve\rightarrow 0 } \int_{0}^\infty \frac{\mrmd t}{t}\, e^{\pm i t (u \pm i \ve) } \,\Tr\, U^{\pm}(t),
\end{equation}
where $U^{\pm}(t) = e^{\mp i \msfH t}$ is the propagator\footnote{For simplicity in future calculations, the $\frac{1}{\hbar}$ factor in the exponential is canceled by scaling $t \rightarrow \hbar t$. }, which governs the flow generated by $\msfH$. It is also possible to incorporate the analytical continuations into integration contours,
\begin{equation}\label{Action_AnalyticalContinuation3}
\G^\pm(u) = - \lim_{\ve \rightarrow 0 }\, \int_{0\pm i \ve}^{\infty \pm i\ve} \frac{\mrmd t}{t}\, e^{\pm i t u }\, \Tr\, U^{\pm}(t) .
\end{equation} 
In this form, the spectral information arises from the singularities on the complex $t$-plane, which are intimately related to periods of classical orbits \cite{voros94}. Note that the integrand in \eqref{Action_AnalyticalContinuation3} is already singular at $t=0$, which corresponds to stationary classical motion. In Section \ref{Section: AHO}, for  quantum anharmonic oscillators, we will explain how the spectral information for a perturbative sector, which is related to the stationary classical motion, emerges from this singularity. First we will continue our discussion with the perturbative expansion of $\Tr U^\pm$ and its recursive structure.
\section{Expansion in $D$ Dimensions}\label{Section: ExpansionArbitraryDimension}
Before describing our recursive scheme for the perturbative expansion of $\Tr\, U^\pm(t)$, let's first investigate its general perturbative structure for a Hermitian operator $\msfH$ given in the following form:
\begin{equation}\label{OperatorGeneralForm}
\msfH(\mbfp,\mbfx)  = \msfT(\mbfp) + \msfV(\mbfx),
\end{equation}
where $\msfT$ and $\msfV$ are operator valued functions of  appropriately chosen canonical variables $\mbfx$ and $\mbfp$ such that they form a $2D$ dimensional phase space. From the quantum mechanical point of view, $\msfH$ can be considered as a generalized Hamiltonian. Moreover, from ordinary QM, we know that projecting $\msfH$ onto $\mbfx$-space, the operator $\mbfp$ starts acting as a derivative operator and vice versa. From this fact, one can easily deduce 
\begin{equation}\label{CommutationDerivative}
[\mbfp, \msfV(\mbfx)] = -i\hbar \nabla_\mbfx \msfV(\mbfx) \qquad , \qquad [\mbfx,\msfT(\mbfp)] = i \hbar \nabla_\mbfp \msfT(\mbfp), 
\end{equation}
and the well-known commutation relation between the canonical variables,
\begin{equation}\label{CommutationRelation_XP}
[x^i,p^j] = i \hbar\, \d^{ij} .
\end{equation} 
The general structure of the perturbative expansion of $U^\pm$ can be examined by using the Zassenhauss formula,
\begin{equation}\label{Propagator_Zassenhauss}
U^\pm(t) = e^{\mp i \msfT(\mbfp) t}\, e^{\mp i t \msfV(\mbfx)}\, e^{\pm \frac{t^2}{4}\left[\msfT(\mbfp), \msfV(\mbfx)  \right] }\, e^{\pm \frac{i t^3}{3!}\big(2[\msfV(\mbfx),[\msfT(\mbfp),\msfV(\mbfx)]] + [\msfT(\mbfp),[\msfT(\mbfp),\msfV(\mbfx)]]] \big)} \dots
\end{equation}
Together with \eqref{CommutationDerivative}, it is easy to see that the sequence of exponents in \eqref{Propagator_Zassenhauss} correspond to a \textit{derivative} expansion. Besides this, expanding these exponentials, we can get another expansion which we call \textit{coupling} expansion. This simple observation shows that a perturbative analysis of $U^\pm(t)$ with an operator $\msfH$ as in \eqref{OperatorGeneralForm} should be treated as a double expansion.   \\
\\
Despite the simplicity of the discussion above, the Zassenhauss formula is not a convenient way for practical calculations. Instead, we take a step back and re-write the propagator 
as a time ordered exponential,
\begin{equation}\label{Propagator_TimeOrderedExponential}
U^\pm(t)  = \mcalT \exp\left\{\mp i \int_0^t \mrmd t'\,  \msfH(\mbfp,\mbfx) \right\} ,
\end{equation}
which simplifies to an ordinary one when $\msfH$ is $t$ independent. Note that \eqref{Propagator_TimeOrderedExponential} is the solution of 
\begin{equation}\label{Schrodinger_TimeDependent}
\pm i\frac{\mrmd U^\pm(t)}{\mrmd t} = \msfH(\mbfp,\mbfx) U^\pm(t).
\end{equation}
One way to compute \eqref{Propagator_TimeOrderedExponential} is by introducing a Fourier transformation  between the canonical variables and eliminate one of them \cite{Schwinger:1951nm,DeWitt:1975ys,avramidi2019heat}. In the following, we will present a perturbative expansion for $\Tr U^\pm$ inspired by this approach. However, instead of eliminating one of the variables, we will work on the phase space and integrate out $\mbfx$ and $\mbfp$ after computing the perturbative expansion. This approach was initiated in \cite{Moss:1993rc,Moss_1999} however, in these papers, the recursive structure behind the expansion of the time-ordered exponential \eqref{Propagator_TimeOrderedExponential} and its relation to the semi-classical expansion were not mentioned. \\
\\
Let us start by separating the $\msfT(\mbfp)$ part as
\begin{equation}\label{InteractionPicture_1}
U^{\pm}(t) = e^{\mp i t \msfT(\mbfp)} \tU^\pm(t)
\end{equation}
and re-write \eqref{Schrodinger_TimeDependent} as 
\begin{equation}
\pm i \frac{\mrmd \tU^\pm(t)}{\mrmd t} = \msfV_I^\pm \tU^\pm(t),
\end{equation}
where we introduced the interaction picture potential,
\[\msfV_I^\pm = e^{\mp i t \msfT(\mbfp)}\, \msfV(\mbfx)\, e^{\pm i t \msfT(\mbfp)}.  \] With these definitions, $U^\pm(t)$ is expressed as 
\begin{align}
U^\pm(t) &= e^{\mp i \msfT(\mbfp) t}\, \mcalT \exp\left\{\mp i  \int_{0}^t \mrmd t'\, \msfV_I^\pm \right\} \nonumber \\
 & = e^{\mp i \msfT(\mbfp) t}\,\sum_{n=0}\frac{\left(\mp i \right)^n}{n!}  \int_{0}^{t}\prod_{i=1}^n \mrmd t_i\, \mcalT\,\Big\{\msfV^\pm_I(t_1)\dots \msfV^\pm_I(t_n)\Big\}.  
\end{align}
The next step is projecting the operator valued functions onto a $2D$ dimensional phase space using\footnote{See Appendix \ref{Section: Notation} for conventions.}
\begin{equation}
\msfV(\mbfx) |\mbfx\> = V(\mbfx)|\mbfx\> \quad , \quad \msfT(\mbfp)|\mbfp\> = T(\mbfp)|\mbfp\>
\end{equation} 
and  re-writing 
\begin{equation}
\Tr U^\pm(t) =\int\frac{\mrmd^D x\,\mrmd^D p}{(2\pi\hbar)^D}\, e^{\mp i T(\mbfp)t} \sum_{n=0}\frac{\left(\mp i \right)^n}{n!}  \int_{0}^{t}\prod_{i=1}^n \mrmd t_i\, \<\mbfx|\mbfp\> \<\mbfp|\mcalT\,[\msfV^\pm_I(t_1)\dots \msfV^\pm_I(t_n)]|\mbfx\> .  
\end{equation}
This allows us to exchange the commutators with a derivative expansion. In order to do this, we insert an identity operator for each $\msfV_I(t_i)$,
\begin{align}
\<\mbfx|\mbfp\> \<\mbfp|  \msfV^\pm_I(t_i) &= \<\mbfx|\mbfp\> e^{\pm i \msfT(\mbfp) t_i} \int\mrmd^D x'\, \<\mbfp|\mbfx'\>\<\mbfx'|\msfV\, e^{\mp i \msfT(\mbfp)t_i} \nonumber \\ 
& =  e^{\pm i \msfT(\mbfp) t_i} \int \frac{\mrmd^D x'}{(2\pi\hbar)^D} \, e^{- \frac{i\mbfp \cdot (\mbfx'-\mbfx)} {\hbar}} V(\mbfx')\<\mbfx'|\,e^{\pm i \msfT(\mbfp)t_i}. \label{IdentityInsertion_V_I}
\end{align}
At this point, instead inserting a second identity operator for $e^{\pm i \msfT(\mbfp) t_i}$, we expand $V(\mbfx')$ around $\mbfx'=\mbfx$,
\begin{equation}
V(\mbfx')  = \sum_{k=0} \frac{1}{k!}  \Big((\mbfx' - \mbfx)\cdot\nabla_\mbfx \Big)^k\, V(\mbfx).
\end{equation}
This enables us to take $x'$ integral using integration by parts. Then, removing the part we introduced as identity element, we get
\begin{equation}
\<\mbfx|\mbfp\> \<\mbfp| \msfV_I(t_i) = \sum_{k=0}\hbar^k\, W_k^\pm \<\mbfx|\mbfp\> \<\mbfp| ,
\end{equation}
where 
\begin{equation}\label{DerivativeExpansion_Operators}
W^\pm_k = \frac{V^{(k)}(\mbfx)}{k!}\, \msfb_\pm^k (\mbfp,\nabla_\mbfp, t_i) \quad , \quad \msfb_\pm(\mbfp,\nabla_\mbfp, t_i)=i \nabla_\mbfp  \pm \nabla_\mbfp T(\mbfp) t_i \mp \nabla_\mbfp T(\mbfp) t .
\end{equation}
Finally, we can express $\Tr\, U^\pm$ as a time-ordered exponential,
\begin{equation}\label{DerivativeExpansion_TimeOrderedExponential}
\Tr\, U^\pm(t) = \int \frac{\mrmd^D x}{(2\pi\hbar)^D }\, \left\< \mcalT \, \exp \left\{ \mp i \int_0^t \mrmd t' \, \sum_{k=0}\hbar^k \, W_k^\pm  \right\}\right\>_\pm ,
\end{equation}
where \[\< \dots \>_\pm = \int  \mrmd^D p \dots e^{\mp i T(\mbfp) t} . \]

\subsection{Recursion Relation}

Equation \eqref{DerivativeExpansion_TimeOrderedExponential} is still impractical for perturbative calculations. In addition to that, depending on the functions $V(\mbfx)$ and $T(\mbfp)$, the volume integrals might lead to infinities which have no physical implications and they are handled by a normalization technique. In the following, we propose a method that can be used for practical calculations and only infinities we encounter will be related to the physical spectrum. We will extract this physical information without a need to normalize.\\
\\
We start by making use of the time-ordered exponential in \eqref{DerivativeExpansion_TimeOrderedExponential}. It enables us to re-write \eqref{Schrodinger_TimeDependent} as
\begin{equation}\label{Schrodinger_TimeDependent2}
 \pm i \frac{\mrmd \tU^\pm(t)}{\mrmd t} = \sum_{k=0} \hbar^k W_k^\pm \tU^\pm(t) .
\end{equation}
Let us write $\tU$ in an $\hbar$ expansion as well \[ \tU^\pm(t) = \sum_l \tU_l^\pm(t) \hbar^l . \] Then, matching orders in \eqref{Schrodinger_TimeDependent2}, we get
\begin{equation}\label{SchrodingerDerivativeExpansion}
\pm i \frac{\mrmd \tU_m^\pm (t)}{\mrmd t}  = \sum_{l=0}^mW^\pm_l \tU^\pm_{m-l}(t) . 
\end{equation} 
At order $m=0$, the solution is 
\begin{equation}\label{Kernel_LeadingOrder_Derivative}
\tU^\pm_0(t) = \mcalT \exp\left\{\mp i  \int_{0}^{t} \mrmd t'\, W^\pm_0(t')  \right\} = e^{\mp i Vt} .
\end{equation}
For $m\geq 1$, after multiplying \eqref{SchrodingerDerivativeExpansion} with $\left(\tU_0^\pm \right)^{-1}$, we get
\begin{equation}\label{RecursivePropagator}
\tU^\pm_m(t) = \mp i \,\tU^\pm_0(t)\int_{0}^{t}\mrmd t'\, (\tU^\pm_0)^{-1}(t') R^\pm_m(t') , 
\end{equation}
where \[R^\pm_m(t) = \sum_{l=1}^{m}W^\pm_l(t) \tU^\pm_{m-l}(t) .\] Note that each $\tU_m^\pm(t)$ is written in terms of $\tU_{l\leq m}^\pm(t)$. This makes the recursive behaviour of the perturbative expansion evident. To utilize this recursive behaviour, we express $\tU_m$ in terms of $\tU_0$ and $W_l$ only as
\begin{equation}
\tU_m(t) = \tU_0^\pm(t)\sum_{k=1}\tU_{m,k}(t) ,
\end{equation} 
where 
\begin{equation}\label{Propagator_SubDerivaativeExpansion}
  \tU^\pm_{m,k}(t)  =\sum_{\substack{\a_1, \dots , \a_k=1\\  (\a_1 + \dots \a_k = m)}}^{m} (\mp i )^k\int_0^t \mrmd t_1 \int_{0}^{t_1}\mrmd t_2 \dots \int_{0}^{t_{k-1}}\mrmd t_k\,  W_{\a_1}^\pm (t_1) \dots W_{\a_k}^\pm(t_k) .
\end{equation}
In \eqref{Propagator_SubDerivaativeExpansion}, we have used \[(\tU_0^\pm)^{-1} W_\a^\pm \tU_0^\pm = W_\a^\pm , \] which was possible since $U_0^\pm = e^{\mp i t V}$ is $\mbfp$ independent. Finally, let us define the sum of products as
\begin{equation}\label{ProductOfSums_Operators}
Q_{m,k}^\pm = \sum_{\substack{\a_1, \dots , \a_k=1\\  (\a_1 + \dots \a_k = m)}}^{m} W_{\a_1}^\pm (t_1) \dots W_{\a_k}^\pm(t_k).
\end{equation}
For each $m$ and $k$, $Q_{m,k}^\pm$ can be written as a product of two lower order terms. For example, let us separate $W_{\a_1}$ from the rest. Then, we get
\begin{align}
Q^\pm_{m,k} &= \sum_{\a_1}W^\pm_{\a_1} Q^\pm_{m-\a_1, k-1} \nonumber  \\
& = W_1^\pm Q^\pm_{m-1,k-1} + W_2^{\pm}(t_1)Q^\pm_{m-2,k-1} + \dots + W^\pm_{m-(k-1)}(t_1) Q^\pm_{k-1,k-1} \nonumber \\
& = \sum_{l=1}^{m-k+1}W^\pm_l Q^\pm_{m-l,k-1} = \sum_{l=1}^{m-k+1}Q_{l,1}^\pm Q_{m-l,k-1}^\pm , \label{RecursionRelation}
\end{align}
where we set $Q_{\a,0} = \d_{\a,0}$. This is the recursion relation we were looking for. Using this recursion relation, we can explicitly express \eqref{Propagator_SubDerivaativeExpansion} as
\begin{align}
\tU_{m,k}(t) &= \sum_{l=1}^{m-k+1} \int_{0}^t \mrmd t_1\, W_l(t_1)\, \tU_{m-l,k-1}(t_1)\label{RecursionRelation_1} \\
& =\sum_{l=1}^{m-k+1}\frac{ V^{(l)}(\mbfx)}{l!} \int_{0}^t \mrmd t_1\, \,  \msfb^l(\mbfp,\nabla\mbfp,t_{1})\tU_{m-l,k-1}(t_1) .
\end{align} 
Assuming $\tU_{m-l,k-1}$ is already computed, for each $\tU_{m,k}$, we only need to compute one $l^\th $ order differentiation and one $t_k$ integral. This is a crucial point to speed up practical calculations. \\
\\
Combining all of these, we write the actions in an $\hbar$ expansion,
\begin{align}\label{QuantumAction_WKB_Expansion}
\G^\pm (u) & = -\lim_{\ve \rightarrow 0 }\sum_{m=0} \hbar^{m} \int_{0\pm i \ve}^{\infty \pm i \ve} \frac{\mrmd t}{t} e^{\pm iut} \int  \frac{\mrmd^D x}{(2\pi\hbar)^D }\, e^{\mp i V(\mbfx) t}  \, \left\< \sum_{k=1}^{m} \tU^\pm_{m,k}(t) \right\>_\pm.
\end{align}
Note that in $\tU_{m,k}$, $k$  represents the number of potentials, i.e. the order of coupling expansion, while $m$ represents the total number of derivatives acting on these potentials. In our arrangement, at any order $k\leq m$. Higher order terms in the coupling expansion comes from the expansion of $\tU_0^\pm = e^{\mp itV}$, if the $x$ integral could not be taken directly. \\
\\
In addition to generating additional terms for the coupling expansion, $\tU_0^\pm = e^{\pm it V}$ in \eqref{QuantumAction_WKB_Expansion} also enables us to obtain finite results for the $x$ integration as long as we compute it around a minimum of $V(\mbfx)$. For example, in Section \ref{Section: AHO}, we will compute the expansion for quantum anharmonic oscillators around their harmonic minima. In those cases, the $x$ integrals will be Gaussian and with a proper analytical continuation of in complex $t$ plane, they lead to finite results. However, this would not be possible if we use the time-ordered exponential in \eqref{DerivativeExpansion_TimeOrderedExponential} directly. Note also that due to the separation in \eqref{InteractionPicture_1}, the $p$ integrals do not need a further treatment to prevent non-physical infinities.
\subsubsection*{Some Remarks} 
\begin{itemize}
	\item Instead of the definition in \eqref{InteractionPicture_1}, we can split the original propagator as
	\begin{equation}\label{InteractionPicture_2}
		U^\pm(t) = e^{\mp i V(\mbfx) t} \tU^\pm(t) .
	\end{equation}
	This is an equivalent definition and the only difference would be the roles of $\msfT(\mbfp)$ and $\msfV(\mbfx)$ in the double expansion. Following the same procedure, we get the following recursion relation
	\begin{align}\label{RecursionRelation_2}
	P^\pm_{n,k}(t) = \sum_{l=1}^{m-k+1} \int_{0}^t \mrmd t_1\, R_l(t_1)\, P_{n-l,k-1}(t_k),
	\end{align}
	where 
	\begin{equation}\label{Propagator_SubDerivaativeExpansion2}
	P^\pm_{n,k}(t)  =\sum_{\substack{\a_1, \dots , \a_k=1\\  (\a_1 + \dots \a_k = n)}}^{n} (\mp i )^k\int_0^t \mrmd t_1 \int_{0}^{t_1}\mrmd t_2 \dots \int_{0}^{t_{k-1}}\mrmd t_k\,  R_{\a_1}^\pm (t_1) \dots R_{\a_k}^\pm(t_k)
	\end{equation}
	and
	\begin{equation} 
	R^\pm_k = \frac{T^{(k)}(\mbfp)}{k!}\, \msfc_\pm^k (\mbfx,\nabla_\mbfx, t_i) \quad , \quad \msfc_\pm(\mbfx,\nabla_\mbfx, t_i)=i \nabla_\mbfx  \pm  \nabla_\mbfx V(\mbfx) t_i \mp \nabla_\mbfx V(\mbfx) t .
	\end{equation}
	In this case, the actions $\G^\pm$ becomes
	\begin{equation}\label{QuantumAction_WKB_Expansion2}
		\G^\pm (u) = -\lim_{\ve\rightarrow 0 }\sum_{n=0} \hbar^{n} \int_{0\pm i \ve}^{\infty \pm i \ve} \frac{\mrmd t}{t} e^{\pm iut} \int  \frac{\mrmd^D p}{(2\pi\hbar)^D }\, e^{\mp i T(\mbfp) t}  \, \left\< \sum_{k=1}^{n} P^\pm_{n,k}(t) \right\>_\pm ,
	\end{equation}
	where \[\< \dots \>_\pm = \int  \mrmd^D x \dots e^{\mp i V(\mbfx) t}. \]
	
	\item In both of \eqref{QuantumAction_WKB_Expansion} and \eqref{QuantumAction_WKB_Expansion2}, the order $\hbar$ counts the number of derivatives. But the difference is in the first one, it is the number of derivatives on $V(\mbfx)$, while in the latter, it is the one on $T(\mbfp)$. This difference indicates that \eqref{QuantumAction_WKB_Expansion} and \eqref{QuantumAction_WKB_Expansion2} are different expressions of same object. However, as long as we do not truncate one of these expansions, results coming from both approaches would be equal. 
	
	\item The recursion relation \eqref{RecursionRelation} is in the same form with the well-known WKB recursion relation \cite{bender1999advanced}. However, as stated before, since we consider the contributions of branch points through the $t$ integral, we will take $x$ and $p$ integrals directly and this will be our ticket to go to higher dimensions. For completeness, we will also show the equivalence between our method and the standard WKB in one dimension in Appendix \ref{Section: WKB=Derivative}.  
	\item Finally, note that the recursive behaviour is an intrinsic property of the iterated integrals, which stem from the time-ordered exponential, and it is independent of the functions $T(\mbfp)$ and $V(\mbfx)$. This indicates the topological nature of the derivative expansions, and it is totally consistent with the conjectured equivalence of topological recursion and WKB expansion \cite{Eynard:2007kz,Norbury:2015lcn}. 
\end{itemize}
\section{An Example: Anharmonic Oscillators in  Quantum Mechanics}\label{Section: AHO}
Up to this point, we have constructed a recursive expansion formula for the quantum action $\G$ and expressed each term by a number of integrals. In this section, as an illustrative example, we will demonstrate how the spectral information of the $D$ dimensional quantum anharmonic oscillator is obtained. \\
\\
Let us start with setting
\begin{equation}
T(\mbfp) = \frac{\mbfp^2}{2} 
\end{equation}
and 
\begin{equation}
V(\mbfx) = \frac{\mbfx^2}{2} + \l v(\mbfx) ,
\end{equation}
where $v(\mbfx)$ is a higher degree polynomial. Then, the quantum actions in \eqref{QuantumAction_WKB_Expansion} are written as
\begin{align*}\label{QuantumAction_AHO}
\G^\pm (u) & = -\lim_{\ve\rightarrow 0 }\sum_{m=0} \hbar^{m} \int_{0\pm i \ve}^{\infty \pm i \ve} \frac{\mrmd t}{t} e^{\pm iut} \int  \frac{\mrmd^D x\, \mrmd^D p}{(2\pi\hbar)^D }\, e^{\mp i \left(\frac{\mbfx^2}{2} + \l v(\mbfx)\right) t}  \,   \sum_{k=1}^{m} \tU^\pm_{m,k}(t) \, e^{\mp \frac{i \mbfp^2t}{2}} .
\end{align*}
We carried out the computations in three separate stages:
\begin{enumerate}
	\item \textbf{Iterated Integrals:} \label{STEP: IteratedIntegrals}
	We first need to compute the iterated time integrals using the recursion relation in \eqref{RecursionRelation_1}. 
	\begin{itemize} 
		\item In these computations, the operator, \[\msfb_\pm(\mbfp,\nabla_\mbfp, t_i)=i \nabla_\mbfp  \pm \mbfp\, t_i \mp \mbfp\, t  \] serves as a generator of polynomials in $p$ by acting on both the polynomials generated in the lower orders and $e^{\mp  \frac{i\mbfp^2t}{2}}$. We carried out this procedure by using the \texttt{Nest} function in \textit{Mathematica}.
		\item Note that the $\mbfx$ dependent functions are not affected by this procedure. Their multiplication leads to the polynomials in $\mbfx$.
		\item The $t_i$ terms in $\msfb_\pm$ also form polynomials. They can easily be integrated at each order. Note that they will also contribute to the next order in the iteration.
	\end{itemize}
	
	\item \textbf{Phase Space Integrals:} At this point, each term in the expansion is written as a polynomial of $\mbfp$, $\mbfx$ and $t$. 
	\begin{itemize}
		\item Note that $e^{\mp \frac{i \mbfp^2 t}{2}}$ is already in Gaussian form and $e^{\mp i V(\mbfx) t}$ can be made Gaussian by expanding it for small $\l$. Then, we can integrate out $\mbfx$ and $\mbfp$ using the standard Gaussian integrals,
		\begin{align}
		I_{2n} &= \int\mrmd^D z\, e^{\mp\frac{ i t}{2}\sum_{k=1}^D z_k^2} \,  z_{1}^{2n_1}\dots z_{D}^{2n_D} \nonumber \\
		&= \prod_{k=1}^D \int\mrmd z_k \, e^{\mp \frac{i t}{2} z_k^2}\, z_k^{2n_k} = \left(\frac{1}{\pm 2it}\right)^{n} \prod_{k=1}^D \frac{(2 n_k)!}{ n_k!} \sqrt{\frac{2\pi}{\pm i t}}, \label{GaussianIntegral_ArbitraryDimension}
		\end{align}
		where we set $n_1 + \dots n_D = n$ and in order to prevent divergences in the $z_k$ integrals, the analytical continuation of $t$ in appropriate directions is assumed.
		\item For example, at the leading order in the derivative expansion, we get
		\begin{align}
		\G^\pm_{m=0} (u,\l) & = -\lim_{\ve\rightarrow 0 } \int_{0\pm i\ve}^{\infty \pm i \ve} \frac{\mrmd t}{t} e^{\pm iut} \left(\frac{2\pi}{\pm i t}\right)^{\frac{D}{2}} \int  \frac{\mrmd^D x}{(2\pi\hbar)^D }\, e^{\mp \frac{ i \mbfx^2 t}{2}}  \sum_{k=0} \frac{(\mp i \l t)^k }{k!}v^k(\mbfx) \nonumber\\
		& = -\lim_{\ve\rightarrow 0 } \int_{0\pm i \ve}^{\infty \pm i \ve} \frac{\mrmd t}{t} \frac{e^{\pm iut} }{(\pm i t \hbar)^D} \sum_{n=0}\frac{A_{2n}^{(0)}(\l)}{(\pm i t)^n} , \label{Action_LeadingOrder}
		\end{align} 
		where $A_{2n}^{(0)}(\l)$ is a polynomial of $\l$ originating from the Gaussian integral of the $\mbfx^{2n}$ term of $v^k(\mbfx)$.
		\item For the higher order terms, additional contributions to both $\mbfx$ and $\mbfp$ polynomials come from the recursive procedure in stage \ref{STEP: IteratedIntegrals}. This makes the general expression more complicated but it is still easy to handle by a computer.
		\item Observe that  higher order poles at $t=0$ appear in \eqref{Action_LeadingOrder}. They are critical for us since in our setting they are associated with the spectrum of $\msfH$.
	\end{itemize}  
	\item \textbf{From Singularities to Spectrum:} The singularity at $t=0$ is usually handled by zeta function regularization \cite{voros1987spectral}. However, instead we will show that the basic contour integration techniques together with the $i\ve$ prescription we mentioned in Section \ref{Section: SpectralProblem} are sufficient.
	\begin{itemize}
		\item To explain how we handle these singularities, let us continue with \eqref{Action_LeadingOrder}. We start with introducing a cutoff $\L$ at the lower limit of the $t$ integral. This allows us to express \eqref{Action_LeadingOrder} as
		\begin{equation}\label{Action_LeadingOrder2}
		\G^\pm_{m=0}(u,\l) = - \hbar^{ - D}\sum_{n=0}A_{2n}^{(0)}(\l) \,  (-u)^{D+n}\, \lim_{\substack{\ve \rightarrow 0 \\ \L \rightarrow 0}} \frac{E_{D+n+1}(\L \pm i\ve)}{(\L \pm i\ve)^{D+n}} ,
		\end{equation}
		where we used the generalized exponential integral \cite{NIST:DLMF},
		\begin{equation}\label{ExponentialIntegral_Generalized}
		E_\a(z) = z^{\a-1} \int_{z}^{\infty} \mrmd t\, \frac{e^{-t}}{t^\a} .
		\end{equation} 
		
		\item For $\a \in \mathbb{N}$, it can be expressed as
		\begin{equation}\label{ExponentialIntegral_Relations}
		E_\a(z) = \frac{(-z)^{\a-1}}{(\a-1)!} E_1(z) + \frac{e^{-z}}{(\a-1)!} \sum_{k=0}^{\a-2} (\a-k-2)! (-z)^k.
		\end{equation} 
		The second part of \eqref{ExponentialIntegral_Relations} is regular at $z=0 $, while the first part has a branch cut due to the branch points of $E_1(z)$ at $z=0$ and $z=\infty$. This branch cut leads to
		\begin{equation}\label{ExponentialIntegral_Gap}
		E_1(z e^{2 m \pi i}) - E_1(z) = - 2m \pi i \qquad ; \qquad m\in \mathbb{Z}.
		\end{equation}
		Note that to use \eqref{ExponentialIntegral_Gap} in \eqref{Action_LeadingOrder2}, we interpret the analytical continuation as \[ \L - i\ve = (\L + i \ve)e^{2\pi i}. \] Then, at the leading order we have
		\begin{equation}\label{Gap_LeadingOrder}
		\D \G_{0}(u,\l) = \frac{ 2\pi i}{ \hbar^{ D}} \sum_{n=0} \frac{A_{2n}^{(0)}(\l) u^{D+n}}{(D+n)!}.
		\end{equation} 
		
		\item Same technique can be applied to the higher orders and $\D\G(u)$ can be expressed as
		\begin{equation}\label{Gap_AllOrderExpansion}
		\D\G(u,\l,\hbar) = \sum_{m=0}^{\infty} \D \G_{2m}(u,\l) \hbar^{2m} ,
		\end{equation}
		where each $\D\G_{2m}(u,\l)$ corresponds to a series in $u$ and $\l$. In Appendix \ref{Section: NumericalResults}, we will provide numerical results for several anharmonic oscillators in $D=1,2,3$ dimensions.
	\end{itemize}
\end{enumerate}
\subsubsection*{A Comment on Quantization Conditions}
Setting $\l=1$ for convenience, \eqref{Gap_AllOrderExpansion} allows us to express $\D\G$ as a series in $u$ and $\hbar$. However, as we reviewed in Section~\ref{Section: SpectralProblem}, the original spectral quantity is $u$ itself, so it is natural to investigate a method for obtaining a perturbative series for $u$ starting from $\D\G$. In one dimension, this can be achieved by imposing the Bohr-Sommerfeld quantization condition for (an)harmonic oscillators. In our formulation, it is expressed as
\begin{equation}\label{Quantization_OneDimension}
\D\G(u,\hbar) = 2\pi i \left(N + \frac{1}{2}\right) .
\end{equation}
For a simple harmonic oscillator in one dimension, we have
\begin{equation}\label{HarmonicOscillator_1D}
\D\G(u,\hbar) =  \frac{2\pi i u}{\hbar} ,
\end{equation}
and it is obvious that \eqref{Quantization_OneDimension} and \eqref{HarmonicOscillator_1D} lead to correct eigenvalues, i.e. $u = \hbar\left(N + \frac{1}{2}\right)$. For anharmonic cases, we will have an infinite series in $u$ for each $\D\G_{2m}$ in \eqref{Gap_AllOrderExpansion}. In these cases, the perturbative expansion of $u$ is achieved by inverting the series in \eqref{Gap_AllOrderExpansion} \cite{PhysRevD.89.041701,PhysRevD.93.065037,Kreshchuk:2018qpf}. On the other hand, it appears that a meaningful generalization of the Bohr-Sommerfeld quantization condition to higher dimensions remains lacking and further investigation is needed. 
\section{Discussion and Outlook} \label{Section: Discussion}
In this paper, we investigated the recursive nature of the derivative expansion of the quantum action and showed how to implement it in practical calculations for quantum anharmonic oscillators in arbitrary dimensions. In quantum mechanics, the semi-classical expansion, which is represented by a derivative expansion in our language, can also be obtained via WKB methods in one dimension, or path integrals in arbitrary dimensions. However, our method has advantages over both methods since perturbative calculations using path integral becomes cumbersome very quickly and the WKB method is only applicable to effectively one dimensional problems.\\
\\
Besides this practical advantage, the method we used separates the spectral information into two distinct parts. The first part, which is identified as the recursion relation and iterated integrals. is universal. It is the same for all quantum mechanical systems. Moreover, despite some differences in the details, we expect the same structure to be present in many-body systems and effective quantum field theories as well. This reveals a general relation between the classical action and its quantum corrections at different orders in a wide range of quantum theories. As we mentioned in the introduction, this relation has been well studied for one dimensional quantum mechanical models via Picard-Fuchs differential equations and the holomorphic anomaly equation \cite{Kreshchuk:2018qpf,Codesido:2017jwp,Codesido_2017,Raman:2020sgw,Fischbach:2018yiu,Mironov_2010,Basar_2017}. Thus, the current paper can also be interpreted as an extension of those methods to higher dimensions and possibly to more complicated theories.\\
\\
On the other hand, the second part, which is identified as the phase-space integrals, depends on the particular system chosen, and therefore it carries information specific to that system. One important piece of information is the divergent large order behaviour of the semi-classical expansion. Although we postpone examining this to  future work, our systematic construction allows an efficeint computation of high orders and allows us to examine its hidden non-perturbative structure.\\
\\
As we have described, the perturbative spectrum gets contribution from the singular part of the small $t$ expansion in \eqref{QuantumAction_WKB_Expansion}. However, each order in the derivative expansion in \eqref{QuantumAction_WKB_Expansion} contains finite contributions as well. Note that their order by order integration leads to a divergent series (see \cite{Dunne:1999uy} for an example in the 1 loop effective action in QFT). Although, this was not important in our construction one could, before taking $t$ integral, obtain a function by summing the finite part and it would be interesting to examine its contribution to the non-perturbative sector of the spectrum through $t$ integration. \\
\\
Finally, let us finish with some apparent downsides of the method we proposed. The first is the lack of expansions related to non-perturbative sector. In WKB related approaches, these expansions are obtained by integrating along classically non-allowed paths, but we get the spectral information from the singularity at $t=0$ plane and so no non-perturbative term emerges in our calculations. However, as the resurgence theory indicates there should be intimate connection between perturbative and non-perturbative sectors. For genus one potentials this  is described by Matone's relation \cite{Matone:1995rx,GORSKY201533,Basar_2017}. Adapting this to our formalism could be useful to understand the emergence of non-perturbative terms and it can be used to verify the connection between perturbative and non-perturbative sectors in more complicated theories.
\section*{Acknowledgements}
The author is grateful to Dieter Van den Bleeken for detailed discussions on the subject and comments on the manuscript. He also would like to thank Can Kozçaz for various discussions. The author is supported by the Boğaziçi University Research Fund under grant number 20B03D1.

\appendix 
\section{Notations and Conventions} \label{Section: Notation}
\begin{align}
\<\mbfx|\mbfp\> = e^{i\mbfx\cdot\mbfp/\hbar} &\quad \<\mbfp|\mbfx\> =\frac{e^{-i\mbfx\cdot\mbfp/\hbar} }{(2\pi\hbar)^D} 
\end{align}
\begin{align} 
\<\mbfx|\mbfx'\> = \d^D(\mbfx - \mbfx') &\quad \<\mbfp|\mbfp'\> = (2\pi\hbar)^D \d^D(\mbfp - \mbfp') 
\end{align}
\begin{align}
\mathbf{1 } = \int \mrmd^D x \, |\mbfx\>\<\mbfx| = \int \frac{\mrmd^D p}{(2\pi \hbar)^D}|\mbfp\>\<\mbfp|
\end{align}
\begin{equation}
\Tr\, O = \int \mrmd^D x\, \<\mbfx|O|\mbfx\> = \int \frac{\mrmd^D p}{(2\pi \hbar)^D}\, \<\mbfp|O|\mbfp\>
\end{equation}
\section{WKB expansion = Derivative Expansion}\label{Section: WKB=Derivative}
Here, for completeness, we compute the first two non-zero terms in the expansions of the one dimensional non-relativistic quantum mechanics, i.e., $T(p) = \frac{p^2}{2}$, for a general potential $V(x)$. This will show the equivalence between the standard WKB approximation and the derivative expansion in our formalism. In addition to that we will also observe how the \textit{``physical''} singularities transfer from the $t$ integral to the $x$ integral.
\subsubsection*{Leading Order}
At the leading order ($m=0$) in one dimension ($D=1$), \eqref{QuantumAction_WKB_Expansion} simplifies to 
\begin{align}
\G^\pm_0(E) &= -\hbar \lim_{\ve\rightarrow 0 } \int_0^\infty  \frac{\mrmd t}{t} \int \frac{\mrmd x \, \mrmd p }{2\pi\hbar }\, e^{\mp \frac{ip^2t}{2}} e^{\pm i(u \pm i\ve-V)t} .
\end{align}
Performing the $p$ integral, rotating the $t$ integral contour by $e^{\pm\frac{i\pi}{2}}$ and re-scaling $t \rightarrow \frac{t}{u \pm i\ve - V(x)}$, we get
\begin{align}
\G^\pm_0(E) &= e^{\mp \frac{i\pi}{2}}\int_{0}^{\infty} \frac{\mrmd t \, e^{-t}} {\sqrt{2\pi t^3}} \int\mrmd x\, \sqrt{u \pm i\ve -V(x)} .
\end{align}
Note that the $t$ integral is still divergent at the lower boundary. Remark that the branch cut  information is now carried to points giving $u=V(x)$ in $x$ space.  Handling the divergence at $t=0$ by zeta-regularization, we get
\begin{equation}
\D \G (E) = \G^+(E) - \G^-(E) = i\sqrt{2} \sum_{i} \oint_{\a_i} \mrmd x \, \sqrt{u - V(x)} ,
\end{equation}
where the each contour $\a_i$ is taken around the singularities at $u=V(x)$, i.e the turning points. Finally, combining with the quantization condition, we get 
\begin{equation}
\sqrt{2} \oint \mrmd x\, \sqrt{u - V(x)} = 2 \pi \left(N +\frac{1}{2}\right) \hbar ,
\end{equation}
which matches the Bohr-Sommerfeld formula. 

\subsubsection*{Corrections up to $O(\hbar^2)$}
For $m=1$, the action is given as 
\begin{equation}
\G^\pm_{m=1} (E) = -\lim_{\ve\rightarrow 0 }\,  \hbar \int_{0}^{\infty} \frac{\mrmd t}{t} e^{\pm it (u\pm i\ve) } \int \frac{\mrmd x}{2\pi \hbar}\, e^{\mp i V t} \int_{0}^{t} \mrmd t'\, \Big\<   W_1(t') \Big\>_\pm ,
\end{equation}
where 
\[\Big\<W_1\Big\>_\pm = \Big\< V'(x)\, b_\pm(p)\Big\>_\pm =0 .  \]
Thus, at order $\hbar$ there is no contribution to the action.\\
\\
Similarly, for $m=2$, we have
\begin{align*}
\G_{m=2}^\pm(E) &= -\lim_{\ve\rightarrow 0 }\, \hbar^2 \int_{0}^\infty \frac{\mrmd t}{t} e^{\pm it(u\pm i\ve )} \int \mrmd x\, e^{\mp iVt} \nonumber \\
& \left\{\int_{0}^t \mrmd t_1 \frac{V''(x)}{2}\Big\< b_\pm(t_1) b_\pm(t_1) \Big\>_\pm \right.  \left . \mp \int_{0}^t\mrmd t_1\int_{0}^{t_1}\mrmd t_2 \left(V'(x)\right)^2 \Big\< b_\pm(t_1) b\pm(t_2)\Big\>_\pm  \right\} .
\end{align*}
Following the same arguments as for the leading order, we get
\begin{align*}
\D \G_{m=2} &= - \frac{i\hbar^2}{\sqrt{2}}\sum_{i} \oint_{\a_i} \mrmd x \left\{\frac{V''(x)}{24(u - V)^{3/2}}+  \frac{(V'(x))^2 }{32(u - V)^{5/2}}  \right\} \\
& = - i\hbar^2 \frac{\sqrt{2}}{2^6}\sum_i \oint_{\a_i}\frac{(V'(x))^2}{(u - V)^{5/2}} ,
\end{align*}
which reproduces the well known first quantum correction to the WKB approximation.
\section{Results}\label{Section: NumericalResults}
Here we present the results for several anharmonic oscillators in one, two and three dimensions. The computations are done by the implementation of the recursive formula in \eqref{RecursionRelation_1} to \textit{Mathematica}. The results for one dimension match exactly with the ones in the literature \cite{Codesido_2017,Fischbach:2018yiu,PhysRevD.93.065037} and for two and three dimension, to our knowledge, these results appear for the first time.
\paragraph{Cubic Oscillator:} $V(\mbfx) = \frac{\mbfx^2}{2} + \l x_1 \mbfx^2 $ \\
\subparagraph{1 Dimension:}
\begin{align*}
\D \G_0(u) & = u+\frac{15 u^2 \lambda ^2}{4}+\frac{1155 u^3 \lambda ^4}{16}+\frac{255255 u^4 \lambda ^6}{128}+\frac{66927861 u^5 \lambda ^8}{1024}\\
\D \G_2(u) & = \frac{7 \lambda ^2}{16}+\frac{1365 u \lambda ^4}{64}+\frac{285285 u^2 \lambda ^6}{256}+\frac{121246125 u^3 \lambda ^8}{2048}+\frac{51869092275 u^4 \lambda ^{10}}{16384}\\
\D \G_4 (u) &= \frac{119119 \lambda ^6}{2048}+\frac{156165009 u \lambda ^8}{16384}+\frac{67931778915 u^2 \lambda ^{10}}{65536}+\frac{24568660040925 u^3 \lambda ^{12}}{262144}\\ 
\end{align*}
\subparagraph{2 Dimensions:}
\begin{align*}
\D \G_0(u) & = \frac{u^2}{2}+2 u^3 \lambda ^2+30 u^4 \lambda ^4+672 u^5 \lambda ^6+18480 u^6 \lambda ^8\\
\D \G_2(u) & =-\frac{1}{12}+\frac{u \lambda ^2}{3}+10 u^2 \lambda ^4+\frac{1120 u^3 \lambda ^6}{3}+15400 u^4 \lambda ^8\\
\D \G_4 (u) &= \frac{11 \lambda ^4}{105}+\frac{92 u \lambda ^6}{3}+3240 u^2 \lambda ^8+265408 u^3 \lambda ^{10}+19347328 u^4 \lambda ^{12}\\ 
\end{align*}
\subparagraph{3 Dimensions:}
\begin{align*}
\D \G_0(u) & = \frac{u^3}{6}+\frac{35 u^4 \lambda ^2}{48}+\frac{3003 u^5 \lambda ^4}{320}+\frac{46189 u^6 \lambda ^6}{256}+\frac{26558675 u^7 \lambda ^8}{6144}\\
\D \G_2(u) & =-\frac{u}{8}-\frac{5 u^2 \lambda ^2}{96}+\frac{77 u^3 \lambda ^4}{128}+\frac{36465 u^4 \lambda ^6}{1024}+\frac{37182145 u^5 \lambda ^8}{24576} \\
\D \G_4 (u) &= -\frac{2369 \lambda ^2}{80640}-\frac{2869 u \lambda ^4}{5120}-\frac{265551 u^2 \lambda ^6}{28672}+\frac{30808063 u^3 \lambda ^8}{294912}\\ 
\end{align*}
\paragraph{Quartic Oscillator:} $V(\mbfx) = \frac{\mbfx^2}{2} + \l (\mbfx^2)^2 $ \\
\subparagraph{1 Dimension:}
\begin{align*}
\D \G_0(u) & = u-\frac{3 u^2 \lambda }{2}+\frac{35 u^3 \lambda ^2}{4}-\frac{1155 u^4 \lambda ^3}{16}+\frac{45045 u^5 \lambda ^4}{64}-\frac{969969 u^6 \lambda ^5}{128}\\
\D \G_2(u) & =-\frac{3 \lambda}{8}+\frac{85 u \lambda ^2}{16}-\frac{2625 u^2 \lambda ^3}{32}+\frac{165165 u^3 \lambda ^4}{128}-\frac{10465455 u^4 \lambda ^5}{512} \\
\D \G_4 (u) &= -\frac{1995 \lambda ^3}{256} +\frac{400785 u \lambda ^4}{1024}-\frac{26249223 u^2 \lambda ^5}{2048}+\frac{1419711293 u^3 \lambda ^6}{4096}\\ 
\end{align*}
\subparagraph{2 Dimensions:}
\begin{align*}
\D \G_0(u) & = \frac{u^2}{2}-\frac{4 u^3 \lambda }{3}+8 u^4 \lambda ^2-64 u^5 \lambda ^3+\frac{1792 u^6 \lambda ^4}{3}\\
\D \G_2(u) & =-\frac{1}{12}-\frac{2 u \lambda }{3}+8 u^2 \lambda ^2-\frac{320 u^3 \lambda ^3}{3}+\frac{4480 u^4 \lambda ^4}{3}\\
\D \G_4 (u) &=\frac{4 \lambda ^2}{9}-\frac{2752 u \lambda ^3}{105}+\frac{17536 u^2 \lambda ^4}{21}-\frac{192512 u^3 \lambda ^5}{9}+489472 u^4 \lambda ^6 \\ 
\end{align*}
\subparagraph{3 Dimensions:}
\begin{align*}
\D \G_0(u) & = \frac{u^3}{6}-\frac{5 u^4 \lambda }{8}+\frac{63 u^5 \lambda ^2}{16}-\frac{1001 u^6 \lambda ^3}{32}+\frac{36465 u^7 \lambda ^4}{128}\\
\D \G_2(u) & = -\frac{u}{8}-\frac{5 u^2 \lambda }{16}+\frac{455 u^3 \lambda ^2}{96}-\frac{8085 u^4 \lambda ^3}{128}+\frac{435435 u^5 \lambda ^4}{512}\\
\D \G_4 (u) &= \frac{269 \lambda }{4480}+\frac{523 u \lambda ^2}{1280}-\frac{67479 u^2 \lambda ^3}{2560}+\frac{7501923 u^3 \lambda ^4}{10240}-\frac{136473909 u^4 \lambda ^5}{8192}\\ 
\end{align*}
\paragraph{Quintic Oscillator:} $V(\mbfx) = \frac{\mbfx^2}{2} + \l x_1(\mbfx^2)^2 $ \\
\subparagraph{1 Dimension:}
\begin{align*}
\D \G_0(u) & = u+\frac{315 u^4 \lambda ^2}{16}+\frac{692835 u^7 \lambda ^4}{128}+\frac{9704539845 u^{10} \lambda ^6}{4096}+\frac{166966608033225 u^{13} \lambda ^8}{131072}\\
\D \G_2(u) & =\frac{1085 u^2 \lambda ^2}{32}+\frac{15570555 u^5 \lambda ^4}{512}+\frac{456782651325 u^8 \lambda ^6}{16384}+\frac{6734319857340075 u^{11} \lambda ^8}{262144}\\
\D \G_4 (u) &= \frac{1107 \lambda ^2}{256}+\frac{96201105 u^3 \lambda ^4}{2048}+\frac{4140194663605 u^6 \lambda ^6}{32768}+\frac{489884540580510075 u^9 \lambda ^8}{2097152}\\ 
\end{align*}
\subparagraph{2 Dimensions:}
\begin{align*}
\D \G_0(u) & = \frac{u^2}{2}+8 u^5 \lambda ^2+1440 u^8 \lambda ^4+465920 u^{11} \lambda ^6+198451200 u^{14} \lambda ^8\\
\D \G_2(u) & =-\frac{1}{12}+\frac{56 u^3 \lambda ^2}{3}+9408 u^6 \lambda ^4+\frac{17937920 u^9 \lambda ^6}{3}+4213780480 u^{12} \lambda ^8\\
\D \G_4 (u) &= \frac{44 u \lambda ^2}{7}+20704 u^4 \lambda ^4+\frac{235023360 u^7 \lambda ^6}{7}+\frac{222716628992 u^{10} \lambda ^8}{5}\\ 
\end{align*}
\subparagraph{3 Dimensions:}
\begin{align*}
\D \G_0(u) & = \frac{u^3}{6}+\frac{77 u^6 \lambda ^2}{32}+\frac{323323 u^9 \lambda ^4}{1024}+\frac{1302340845 u^{12} \lambda ^6}{16384}+\frac{7244053893505 u^{15} \lambda ^8}{262144}\\
\D \G_2(u) & = -\frac{u}{8}+\frac{805 u^4 \lambda ^2}{128}+\frac{2263261 u^7 \lambda ^4}{1024}+\frac{34659070875 u^{10} \lambda ^6}{32768}+\frac{624578793013175 u^{13} \lambda ^8}{1048576}\\
\D \G_4 (u) &= \frac{83819 u^2 \lambda ^2}{23040}+\frac{1121359525 u^5 \lambda ^4}{172032}+\frac{1891956467895 u^8 \lambda ^6}{262144}+\frac{91473021008360675 u^{11} \lambda ^8}{12582912}\\ 
\end{align*}

\bibliographystyle{JHEP}
\bibliography{Recursive_QA.bib}

\providecommand{\href}[2]{#2}\begingroup\raggedright\begin{thebibliography}{10}

\bibitem{Aniceto:2018bis}
I.~Aniceto, G.~Basar and R.~Schiappa, \emph{{A Primer on Resurgent Transseries
  and Their Asymptotics}},
  \href{https://doi.org/10.1016/j.physrep.2019.02.003}{\emph{Phys. Rept.}
  {\bfseries 809} (2019) 1--135}.

\bibitem{percival1977semiclassical}
I.~C. Percival, \emph{Semiclassical theory of bound states}, {\emph{Adv. Chem.
  Phys} {\bfseries 36} (1977) 61}.

\bibitem{brack1997semiclassical}
M.~Brack and R.~Bhaduri, \emph{Semiclassical Physics}.
\newblock Frontiers in physics. Avalon Publishing, 1997.

\bibitem{Sulejmanpasic:2016fwr}
T.~Sulejmanpasic and M.~\"Unsal, \emph{{Aspects of perturbation theory in
  quantum mechanics: The BenderWu Mathematica \textregistered{} package}},
  \href{https://doi.org/10.1016/j.cpc.2017.11.018}{\emph{Comput. Phys. Commun.}
  {\bfseries 228} (2018) 273--289},
  [\href{https://arxiv.org/abs/1608.08256}{{\ttfamily 1608.08256}}].

\bibitem{Bender:1973rz}
C.~M. Bender and T.~Wu, \emph{{Anharmonic oscillator. 2: A Study of
  perturbation theory in large order}},
  \href{https://doi.org/10.1103/PhysRevD.7.1620}{\emph{Phys. Rev. D} {\bfseries
  7} (1973) 1620--1636}.

\bibitem{Basar_2017}
G.~Basar, G.~V. Dunne and M.~\"Unsal, \emph{Quantum geometry of resurgent
  perturbative/nonperturbative relations},
  \href{https://doi.org/10.1007/jhep05(2017)087}{\emph{Journal of High Energy
  Physics} {\bfseries 2017} (May, 2017) }.

\bibitem{Raman:2020sgw}
M.~Raman and P.~Bala~Subramanian, \emph{{Chebyshev wells: Periods,
  deformations, and resurgence}},
  \href{https://doi.org/10.1103/PhysRevD.101.126014}{\emph{Phys. Rev. D}
  {\bfseries 101} (2020) 126014}.

\bibitem{Mironov_2010}
A.~Mironov and A.~Morozov, \emph{Nekrasov functions and exact bohr-sommerfeld
  integrals}, \href{https://doi.org/10.1007/jhep04(2010)040}{\emph{Journal of
  High Energy Physics} {\bfseries 2010} (Apr, 2010) }.

\bibitem{Kreshchuk:2018qpf}
M.~Kreshchuk and T.~Gulden, \emph{{The Picard--Fuchs equation in classical and
  quantum physics: application to higher-order WKB method}},
  \href{https://doi.org/10.1088/1751-8121/aaf272}{\emph{J. Phys. A} {\bfseries
  52} (2019) 155301}.

\bibitem{Fischbach:2018yiu}
F.~Fischbach, A.~Klemm and C.~Nega, \emph{{WKB Method and Quantum Periods
  beyond Genus One}}, \href{https://doi.org/10.1088/1751-8121/aae8b0}{\emph{J.
  Phys. A} {\bfseries 52} (2019) 075402}.

\bibitem{Codesido_2017}
S.~Codesido and M.~Mariño, \emph{Holomorphic anomaly and quantum mechanics},
  \href{https://doi.org/10.1088/1751-8121/aa9e77}{\emph{Journal of Physics A:
  Mathematical and Theoretical} {\bfseries 51} (2018) }.

\bibitem{Codesido:2017jwp}
S.~Codesido, M.~Marino and R.~Schiappa, \emph{{Non-Perturbative Quantum
  Mechanics from Non-Perturbative Strings}},
  \href{https://doi.org/10.1007/s00023-018-0751-x}{\emph{Annales Henri
  Poincare} {\bfseries 20} (2019) 543--603}.

\bibitem{NEKRASOV_2010}
N.~A. NEKRASOV and S.~L. SHATASHVILI, \emph{Quantization of integrable systems
  and four dimensional gauge theories},
  \href{https://doi.org/10.1142/9789814304634_0015}{\emph{XVIth International
  Congress on Mathematical Physics} (Mar, 2010) }.

\bibitem{Eynard:2007kz}
B.~Eynard and N.~Orantin, \emph{{Invariants of algebraic curves and topological
  expansion}}, \href{https://doi.org/10.4310/CNTP.2007.v1.n2.a4}{\emph{Commun.
  Num. Theor. Phys.} {\bfseries 1} (2007) 347--452}.

\bibitem{Norbury:2015lcn}
P.~Norbury, \emph{{Quantum curves and topological recursion}}, {\emph{Proc.
  Symp. Pure Math.} {\bfseries 93} (2015) 41},
  [\href{https://arxiv.org/abs/1502.04394}{{\ttfamily 1502.04394}}].

\bibitem{DUNNE_2005}
G.~V. Dunne, \emph{Heisenberg-euler effective lagrangians: Basics and
  extensions}, \href{https://doi.org/10.1142/9789812775344_0014}{\emph{From
  Fields to Strings: Circumnavigating Theoretical Physics} (Feb, 2005)
  445–522}.

\bibitem{voros1992}
A.~Voros, \emph{Spectral zeta functions},  in \emph{Zeta Functions in
  Geometry}, (Tokyo, Japan), pp.~327--358, Mathematical Society of Japan, 1992,
  \href{https://doi.org/10.2969/aspm/02110327}{DOI}.

\bibitem{Eden:1966dnq}
R.~J. Eden, P.~V. Landshoff, D.~I. Olive and J.~C. Polkinghorne, \emph{{The
  analytic S-matrix}}.
\newblock Cambridge Univ. Press, Cambridge, 1966.

\bibitem{Zinn_Justin_2004}
J.~Zinn-Justin and U.~D. Jentschura, \emph{Multi-instantons and exact results
  i: conjectures, wkb expansions, and instanton interactions},
  \href{https://doi.org/10.1016/j.aop.2004.04.004}{\emph{Annals of Physics}
  {\bfseries 313} (Sep, 2004) 197–267}.

\bibitem{1970JMP....11.1791G}
M.~C. {Gutzwiller}, \emph{{Energy Spectrum According to Classical Mechanics}},
  \href{https://doi.org/10.1063/1.1665328}{\emph{Journal of Mathematical
  Physics} {\bfseries 11} (June, 1970) 1791--1806}.

\bibitem{1971JMP....12..343G}
M.~C. {Gutzwiller}, \emph{{Periodic Orbits and Classical Quantization
  Conditions}}, \href{https://doi.org/10.1063/1.1665596}{\emph{Journal of
  Mathematical Physics} {\bfseries 12} (Mar., 1971) 343--358}.

\bibitem{voros94}
A.~Voros, \emph{{Aspects of Semiclassical Theory in the Presence of Classical
  Chaos}}, \href{https://doi.org/10.1143/PTP.116.17}{\emph{Progress of
  Theoretical Physics Supplement} {\bfseries 116} (02, 1994) 17--44}.

\bibitem{Schwinger:1951nm}
J.~S. Schwinger, \emph{{On gauge invariance and vacuum polarization}},
  \href{https://doi.org/10.1103/PhysRev.82.664}{\emph{Phys. Rev.} {\bfseries
  82} (1951) 664--679}.

\bibitem{DeWitt:1975ys}
B.~S. DeWitt, \emph{{Quantum Field Theory in Curved Space-Time}},
  \href{https://doi.org/10.1016/0370-1573(75)90051-4}{\emph{Phys. Rept.}
  {\bfseries 19} (1975) 295--357}.

\bibitem{avramidi2019heat}
I.~Avramidi, \emph{Heat Kernel Method and its Applications}.
\newblock Springer International Publishing, 2019.

\bibitem{Moss:1993rc}
I.~Moss and S.~Poletti, \emph{{Finite temperature effective actions for gauge
  fields}}, \href{https://doi.org/10.1103/PhysRevD.47.5477}{\emph{Phys. Rev. D}
  {\bfseries 47} (1993) 5477--5486}.

\bibitem{Moss_1999}
I.~G. Moss and W.~Naylor, \emph{Diagrams for heat kernel expansions},
  \href{https://doi.org/10.1088/0264-9381/16/8/304}{\emph{Classical and Quantum
  Gravity} {\bfseries 16} (Jul, 1999) 2611–2624}.

\bibitem{bender1999advanced}
C.~Bender and S.~Orszag, \emph{Advanced Mathematical Methods for Scientists and
  Engineers I: Asymptotic Methods and Perturbation Theory}.
\newblock Advanced Mathematical Methods for Scientists and Engineers. Springer,
  1999.

\bibitem{voros1987spectral}
A.~Voros, \emph{Spectral functions, special functions and the selberg zeta
  function}, {\emph{Communications in Mathematical Physics} {\bfseries 110}
  (1987) 439--465}.

\bibitem{NIST:DLMF}
``{\it NIST Digital Library of Mathematical Functions}.''
  http://dlmf.nist.gov/, Release 1.0.26 of 2020-03-15.

\bibitem{PhysRevD.89.041701}
G.~V. Dunne and M.~\"Unsal, \emph{Generating nonperturbative physics from
  perturbation theory},
  \href{https://doi.org/10.1103/PhysRevD.89.041701}{\emph{Phys. Rev. D}
  {\bfseries 89} (Feb, 2014) 041701(R)}.

\bibitem{PhysRevD.93.065037}
I.~Gahramanov and K.~Tezgin, \emph{Remark on the dunne-\"unsal relation in
  exact semiclassics},
  \href{https://doi.org/10.1103/PhysRevD.93.065037}{\emph{Phys. Rev. D}
  {\bfseries 93} (Mar, 2016) 065037}.

\bibitem{Dunne:1999uy}
G.~V. Dunne and T.~M. Hall, \emph{{Borel summation of the derivative expansion
  and effective actions}},
  \href{https://doi.org/10.1103/PhysRevD.60.065002}{\emph{Phys. Rev.}
  {\bfseries D60} (1999) 065002}.

\bibitem{Matone:1995rx}
M.~Matone, \emph{{Instantons and recursion relations in N=2 SUSY gauge
  theory}}, \href{https://doi.org/10.1016/0370-2693(95)00920-G}{\emph{Phys.
  Lett. B} {\bfseries 357} (1995) 342--348}.

\bibitem{GORSKY201533}
A.~Gorsky and A.~Milekhin, \emph{Rg-whitham dynamics and complex hamiltonian
  systems},
  \href{https://doi.org/https://doi.org/10.1016/j.nuclphysb.2015.03.028}{\emph{Nuclear
  Physics B} {\bfseries 895} (2015) 33--63}.

\end{thebibliography}\endgroup
\end{document}